\documentclass[11pt]{article}
\usepackage{moriond,epsfig,amsmath,amssymb}

\def\Journal#1#2#3#4{{#1} {\bf #2}, #3 (#4)}

\def\be{\begin{equation}}
\def\ee{\end{equation}}
\def\bea{\begin{eqnarray}}
\def\eea{\end{eqnarray}}

\newcommand{\ol}[1]{\overline{#1}}
\begin{document}
\vspace*{4cm}
\title{SUPERSYMMETRIC MODELS WITH LIGHT HIGGSINOS}

\author{ F.~BR\"UMMER }

\address{Deutsches Elektronen-Synchrotron DESY, Notkestra\ss e 85,\\
D-22607 Hamburg, Germany}

\maketitle\abstracts{In the Minimal Supersymmetric Standard Model, the higgsinos can have masses around the electroweak scale, while the other supersymmetric particles have TeV-scale masses. This happens in models of gauge-mediated SUSY breaking with a high messenger scale, which are motivated from string theory. For particular choices of the messenger field content, multi-TeV squark and gluino masses naturally lead to a much lower electroweak scale, somewhat similar to focus point supersymmetry. They also induce Higgs masses of 124--126 GeV, while making the discovery of supersymmetry at the LHC unlikely. The light higgsinos will be difficult to see at the LHC but may eventually be discovered at a linear collider.
}

\section{Introduction}

When attempting to reconcile the Minimal Supersymmetric Standard Model (MSSM) with the latest LHC results, one is confronted with two major puzzles. Why have no squarks and gluinos been observed yet? And, taking seriously the indications for a 124--126 GeV Higgs, how can the Higgs be so heavy without spoiling naturalness? Furthermore, given that the hopes for an early discovery of supersymmetry were disappointed, the all-important question regarding the ongoing SUSY searches becomes once more: What can we expect to see?

This article presents an attempt to provide answers to these questions. We review models~\cite{Brummer:2011yd,Bobrovskyi:2011jj,Brummer:2012zc} in which soft SUSY breaking terms are generated by an interplay of gauge-mediated and gravity-mediated supersymmetry breaking. This, as we will argue, leads to mass spectra in which all supersymmetric particles can be very heavy, up to multiple TeV. The only exception are two higgsino-like neutralinos and a higgsino-like chargino, whose masses are around the electroweak scale. First-generation squarks and gluinos have evaded detection so far because they are out of reach. Heavy third-generation squarks contribute large radiative corrections to the lightest Higgs mass, enough to increase it to around 125 GeV. Some of the models predict the electroweak symmetry breaking scale to be far below the typical soft mass scale, thus alleviating the naturalness problem in a similar way as focus point supersymmetry. Finally, should our scenario be realized in Nature, supersymmetry will be difficult to find at the LHC. It should however be possible to find the light higgsinos at a future linear collider, and possibly to constrain our models using LHC searches for monojets and missing energy. 

\section{Gauge mediation versus gravity mediation}

In the MSSM there is a single dimensionful parameter which is allowed by unbroken supersymmetry: the higgsino mass $\mu$. All other mass parameters originate from supersymmetry breaking, and are therefore naturally all of the same order of magnitude $m_{\rm soft}$. By contrast, a priori one might expect $\mu$ to be either zero, or of the order of $M_{\rm Planck}$. However, $\mu\lesssim 100$ GeV is excluded by direct chargino searches, and $\mu\gg 1$ TeV considerably increases the fine-tuning required to obtain the proper electroweak scale (and tends to spoil unification). A realistic model of supersymmetry breaking should give a $\mu$ term of the order of the electroweak scale. This is the famous ``$\mu$ problem''.

In gravity-mediated supersymmetry breaking, the SUSY-breaking hidden sector is connected to the MSSM by $M_{\rm Planck}$-suppressed higher-dimensional operators. This induces soft SUSY breaking terms which are generically of the order of the gravitino mass,
\be
m_{\rm soft}\simeq m_{3/2}\,.
\ee
The $\mu$ problem can be solved if an effective $\mu\simeq m_{3/2}$ is induced after SUSY breaking, either from K\"ahler potential terms \cite{Giudice:1988yz} or from superpotential terms \cite{Casas:1992mk}. 

By contrast, in minimal gauge-mediated SUSY breaking \cite{Giudice:1998bp}, soft masses are induced by loops of massive messenger states, which couple to the MSSM only through their gauge charges. The soft terms are of the order
\be
m_{\rm soft}\simeq m_{3/2}\cdot\frac{M_{\rm Planck}}{M_{\rm mess.}}\cdot\frac{1}{16\pi^2}\,.
\ee
This mechanism does not induce a nonzero $\mu$. Additional superpotential couplings between the Higgs fields and the messengers are required to obtain $\mu\neq 0$, but it is difficult to explicitly construct a realistic model in this way, because such additional couplings generically induce a too large Higgs mass mixing parameter $B_\mu$ \cite{Dvali:1996cu}.

It is possible to combine the mechanisms of gravity mediation and gauge mediation, whereupon $\mu$ is induced by gravity mediation alone, while the soft SUSY breaking parameters receive contributions both from gravity mediation and from gauge mediation. Models of hybrid gauge-gravity mediation are not commonly encountered in the literature for several reasons:
\begin{enumerate}
 \item The main advantage of gauge mediation is that it generates flavour-universal soft terms. In gravity mediation, on the other hand, the flavour structure of the soft masses and trilinear parameters is not predicted, and unacceptably large flavour-changing neutral currents can result unless there is some other mechanism ensuring flavour universality.
\item Within gauge-mediated models, gravity-mediated effects are always present, but negligible unless the messenger mass scale is around $M_{\rm mess.}\simeq M_{\rm Planck}/(16\pi^2)$. Comparable effects from gauge and gravity mediation thus require a coincidence of scales which needs to be explained.
 \item Gauge-mediated models are predictive, in the sense that they can be constructed as fully renormalizable theories with fixed particle content and few parameters. Allowing for sizeable gravity-mediated contributions to the soft terms spoils this predictivity by introducing more parameters.
\end{enumerate}

This article is concerned with a particular class of models of hybrid gauge-gravity mediation in which all of these points can be addressed. They are motivated by string-theoretic constructions aiming to obtain the MSSM from heterotic orbifold compactifications or from F-theory. Their most characteristic feature is that they contain, besides the MSSM matter and Higgs fields, also ``exotic matter'' in incomplete vector-like representations of the Grand Unified gauge group. The multiplicities of these exotics can be quite large, ${\cal O}$(few dozen). They obtain masses by coupling to MSSM singlet fields which take non-vanishing expectation values. In a complete model, these singlets would be part of the hidden sector, whereupon the exotics become gauge mediation messengers. This framework naturally leads to hybrid gauge-gravity mediation. As to the above points, we note that
\begin{enumerate}
 \item The SUSY flavour problem must be addressed within the underlying string model. For instance, certain heterotic compactifications exhibit suitable flavour symmetries \cite{Nilles:2012cy}.
 \item The scale $M_{\rm Planck}/(16\pi^2)\approx M_{\rm GUT}$ is a natural scale for the messengers to decouple. It can be related to the volume of the compact internal space in some string model, thus naturally setting the GUT scale in models where the GUT symmetry is broken to the MSSM gauge group by compactification. Its dynamical origin may be traced back \cite{Buchmuller:2005jr} to the appearance of field-dependent Fayet-Iliopoulos terms, whose characteristic size is indeed a loop factor below $M_{\rm Planck}$.
\item Since the messenger multiplicities are large, gauge mediation dominates over gravity mediation. The additional parameters introduced by gravity mediation affect the resulting MSSM spectrum only at the subleading level, except for the $\mu$ parameter which is solely induced by gravity mediation.
\end{enumerate}

The resulting soft mass patterns show some very distinctive features, three of which are especially remarkable. To start with, there is no gaugino mass unification (i.e.~the ratios $M_a/g_a^2$ are not universal, unlike in many other SUSY GUT models), because the messenger fields form incomplete GUT multiplets. Secondly, the ratio $\tan(\beta)$ of Higgs expectation values is large, as a consequence of the  gravity-mediated $B_\mu$ parameter being small. And finally, two higgsino-like neutralinos and a higgsino-like chargino are by far the lightest MSSM states. The reason is that soft SUSY-breaking masses are dominated by gauge-mediated contributions, which are enhanced by large messenger multiplicities. On the other hand, the higgsino mass $\mu$ is induced by gravity mediation, and therefore naturally an order of magnitude smaller. Indeed, the higgsino masses can naturally be of the order of $100$ GeV, with all other superparticles heavier than a TeV.   

\section{An example model}
\begin{table}
\caption{\emph{Left,} the messenger content of a heterotic orbifold
model, with their charges under ${\rm SU}(3)\times{\rm SU}(2)\times{\rm U}(1)$. \emph{Right,} the MSSM mass spectrum of the same model, for a typical choice of parameters. Note that the higgsinos $\chi^0_{1,2}$ and $\chi^\pm_1$ are significantly lighter than the other supersymmetric particles, and nearly degenerate in mass.}\label{messengers} 
\begin{center}
\begin{tabular}{|c|c|c|}\hline field & representation & multiplicity \\ \hline
$d$ & $({\bf 3},{\bf 1})_{-1/3}$ & $4$ \\ 
$\tilde d$ & $(\ol {\bf 3},{\bf 1})_{1/3}$ & $4$ \\ 
$\ell$ & $({\bf 1},{\bf 2})_{1/2}$ & $4$  \\ 
$\tilde\ell$ & $({\bf 1},{\bf 2})_{-1/2}$ & $4$ \\ 
$m$ & $({\bf 1},{\bf 2})_{\,0}$ & $8$ \\ 
$s^+$ & $({\bf 1},{\bf 1})_{1/2}$ & $16$ \\ 
$s^-$ & $({\bf 1},{\bf 1})_{-1/2}$ & $16$ \\ \hline
\end{tabular}$\qquad$
\begin{tabular}{|c|c|c|}\hline particle &  mass [GeV]\\ \hline
$h_0$ & $117$\\ 
$\chi^0_1$ &  $137$ \\ 
$\chi^\pm_1$ & $140$  \\ 
$\chi^0_2$ & $144$ \\ 

$\chi^0_3$ &  $799$ \\ 
$\chi^0_4$ & $1296$ \\ 
$\chi^\pm_2$ & $1296$  \\ 
$H_0$ & $856$ \\ 
$A_0$ & $857$ \\ 
$H^\pm$ & $861$ \\ 
$\tilde g$ & $1453$ \\ 
$\tilde\tau_1$ & $713$ \\
other sleptons & $910 - 1290$\\ 
squarks & $950 - 1750$ \\\hline
\end{tabular}
\end{center}
\end{table}

Table \ref{messengers} contains a sample spectrum of exotic states from a particular heterotic string model \cite{Buchmuller:2005jr}, along with a typical mass spectrum which results from taking these exotics as the messenger sector. The masses were obtained by choosing \cite{Brummer:2011yd} $\mu=A_0=150$ GeV, $B_\mu=(240\text{ GeV})^2$, $m_{3/2}=100$ GeV, and $M_{\rm mess.}=5\cdot 10^{15}$ GeV. A goldstino mixing angle $\phi$, which is also a free parameter in this model, is chosen such that $\tan\phi=1.9$. 

This spectrum, however, is on the brink of being ruled out by direct superparticle searches, and it predicts a Higgs boson which is not compatible with the recent evidence \cite{ATLAS:2012ae,CMS} for a 124--126 GeV Higgs mass. Unless the candidate Higgs signal turns out to be a mere fluctuation, and unless the LHC finds evidence for squarks and gluinos very soon, the above choice of parameters will be ruled out.

While both the Higgs mass and the superpartner masses can be raised by increasing the fundamental SUSY breaking scale, this would, as usual, come at a cost: The fine-tuning required to reproduce the correct electroweak scale increases. This motivates the search for more natural models, which is the topic of the next Section.

\section{Fine-tuning and focus points}\label{focsec}

In the MSSM at large $\tan\beta$ and in the decoupling limit $m_{A^0}\gg m_Z$, the lightest Higgs mass is given by
\begin{equation}\label{higgsmass}
m_{h^0}^2=\,m_Z^2+\frac{3}{4\pi^2}y_t^4 v^2\left(\log\frac{m_{\tilde t}^2}{m_t^2}+\frac{A_t^2}{m_{\tilde t}^2}\left(1-\frac{A_t^2}{12\,m_{\tilde t}^2}\right)\right)+\ldots
\end{equation}
where $m_{\tilde t}$ is the average stop mass and $A_t$ is the trilinear stop mixing parameter. To significantly increase $m_{h^0}$ above $m_Z$, large one-loop corrections from heavy stops are needed. On the other hand, the stop masses feed into the renormalization group equations for the Higgs mass parameters with large coefficients, because of the large top Yukawa coupling. The most natural size of the Higgs mass parameters, and thus of the electroweak symmetry breaking scale, is therefore around $m_{\tilde t}$. A multi-TeV $m_{\tilde t}$ (as needed for a 125 GeV Higgs, in the absence of large stop mixing) while maintaining an EWSB scale around 100 GeV requires considerable fine-tuning; this observation is sometimes called the ``little hierarchy problem''.

\begin{figure}
 \includegraphics[width=77mm]{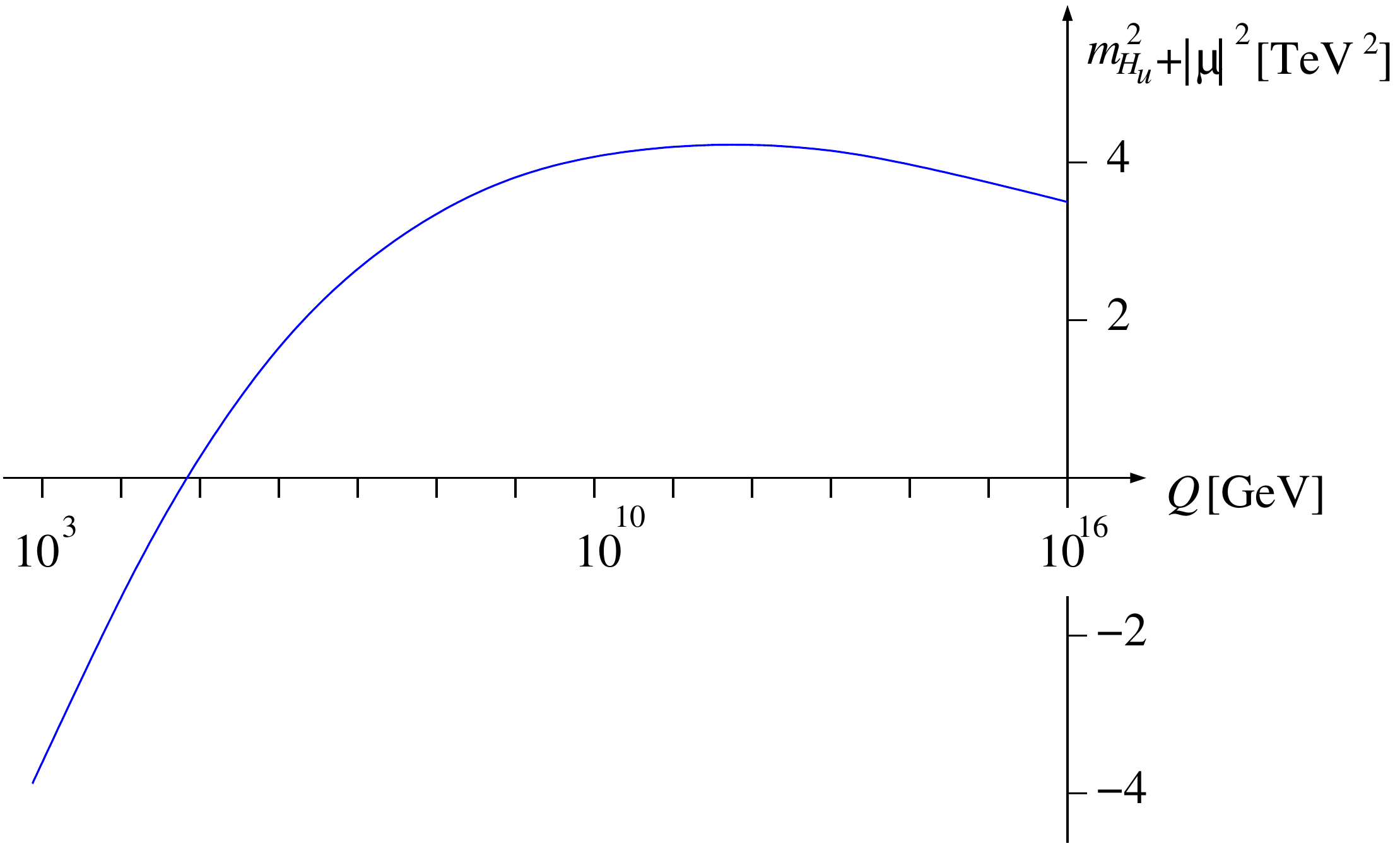}$\quad$
 \includegraphics[width=77mm]{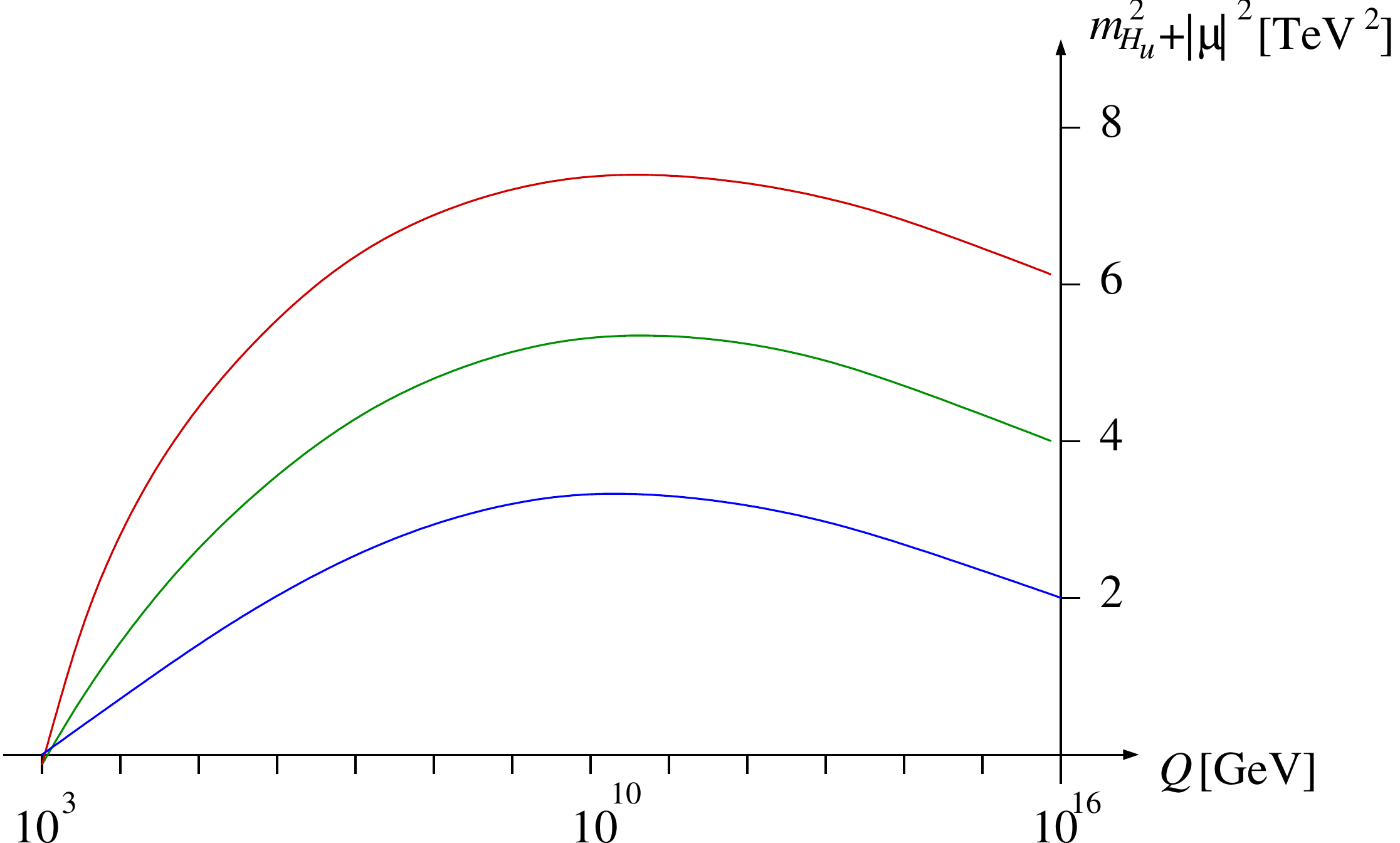}
\caption{Sketch of the RG evolution of $m_{H_u}^2+|\mu|^2$, whose value at the TeV scale determines the electroweak breaking scale according to $m_Z^2=-2\left.(m_{H_u}^2+|\mu|^2)\right|_{\rm TeV}$. {\it Left:} Generically, for TeV-sized soft parameters at the GUT scale, the predicted $m_Z$ would also be ${\cal O}$(TeV). {\it Right:} If the GUT-scale soft masses are subject to suitable relations, the RG trajectories for different GUT-scale values may focus at the TeV scale, and lead to small $m_Z$.} \label{focusfig}
\end{figure}

More precisely, expressing the EWSB order parameter $m_Z$ in terms of the dimensionful GUT-scale parameters, one has
\begin{equation}
\begin{split}\label{nummzsq}
m_Z^2\approx\,&\Bigl(2.25\,M_3^2-0.45\,M_2^2-0.01\,M_1^2+0.19\,M_2 M_3+0.03\,M_1 M_3\\
&+0.74\,m_{\tilde t_R}^2+0.65\,m_{\tilde t_L}^2-0.04\,m_{\tilde b_R}^2-1.32\,m_{H_u}^2-0.09\,m_{H_d}^2\\
&+0.19\,A_0^2-0.40\,A_0 M_3-0.11\,A_0 M_2-0.02\,A_0 M_1
-1.42\,|\mu|^2\left.\Bigr)\right|_{M_{\rm GUT}}
\end{split}
\end{equation}
(here $\tan\beta$ is large, $\tan\beta\approx 50$, and $m_{\tilde t}\approx 1$ TeV). Clearly, if the typical size of the GUT-scale soft terms is $\gg 1$ TeV, then large cancellations are needed to reproduce $m_Z=91$ GeV.

It is an intriguing observation \cite{Chan:1997bi,Feng:1999mn} that for $M_{1,2,3}$ and $\mu$ of the order of the electroweak scale, and for universal GUT-scale scalar soft masses $m_{\tilde t_{L,R}}=m_{\tilde b_R}=m_{H_u}=m_{H_d}$ given by some $m_0$, the coefficients in the second line of Eq.~\eqref{nummzsq} sum up to nearly zero. Therefore, $m_0$ can be several TeV without requiring large cancellations. Pictorially speaking, the RG trajectories of the parameters governing $m_Z$, for various values of $m_0$, ``focus'' close to zero near the electroweak scale. This is sketched in Fig.~\ref{focusfig}.

In our hybrid gauge-gravity mediated models, $m_0$ is not universal, and the gaugino masses are not small. However, a similar focus point can nevertheless appear when also the gaugino masses participate in the cancellation \cite{Horton:2009ed}. In the simplest of our models \cite{Brummer:2012zc}, the gauge-mediated (and thus dominant) contributions to $m_Z$ are determined solely by the messenger content and by the SUSY breaking scale. In a model with $N_3$ pairs of colour triplet messengers and $N_2$ pairs of weak doublet messengers, they are given by
\be\label{delmz}
\Delta m_Z^2\approx\,\bigl(2.25\,N_3^2-0.45\,N_2^2+0.19\,N_2N_3+3.80\,N_3 -1.16\,N_2\bigr)\,m_{\rm GM}^2\,,
\ee
where $m_{\rm GM}\equiv m_{3/2}\cdot\frac{M_{\rm Planck}}{M_{\rm mess.}}\cdot\frac{g^2}{16\pi^2}$ is of the order of the electroweak scale for $m_{3/2}\approx 100$ GeV. The individual soft terms are parametrically larger than $m_{\rm GM}$, because they are enhanced by large messenger numbers $N_{2,3}$: Gaugino masses scale as $N\cdot m_{\rm GM}$, and scalar masses as $\sqrt{N}\cdot m_{\rm GM}$.

In models where $(N_2,N_3)=(23,9)$ or $(28,11)$, the terms on the RHS of Eq.~\eqref{delmz} once again cancel out to great precision. Such models therefore predict an EWSB scale much lower than the typical soft mass scale. They can therefore accommodate a Higgs boson around 125 GeV without sacrificing naturalness, see the mass spectra in Table \ref{tab:spectra}. It would be interesting to find models with these messenger contents in actual string constructions, such as the ones of the ``heterotic mini-landscape'' \cite{Lebedev:2006kn}.

\begin{table}
\caption{Some selected masses in GeV for models with ``focus point'' messenger multiplicities, for typical choices of parameters \protect\cite{Brummer:2012zc}. While the higgsinos $\chi^0_{1,2}$, $\chi^\pm_1$ are light, all other states are likely beyond the reach of the LHC.}\label{tab:spectra}
\begin{center}
\begin{tabular}{|c|c|c|}
\hline
particle & $(N_2,N_3)=(23,9)$ & $(N_2,N_3)=(28,11)$\\ \hline
${h_0}$ &  $123$ &  $124$\\
${\chi^0_1}$ &  $205$ &  $164$\\
${\chi^\pm_1}$ &  $207$ & $166$\\
${\chi^0_2}$ &  $208$ &  $167$\\
${\tilde\tau_1}$ & $1530$ &  $1890$\\ 
${H^0}$ & $1470$ &  $2200$\\
${A}$ & $1480$ &  $2200$\\
${H^\pm}$ & $1480$&  $2200$\\ 
$\chi^0_3$ & $2500$&  $2700$\\
$\chi^0_4$ & $3800$& $4100$\\
$\chi^\pm_2$ & $3800$&  $4100$\\
${\tilde g}$ &  $3800$ &  $4200$\\
${\tilde t_1}$ & $2500$ & $2700$\\ 
${\tilde u_1}$ & $3700$ &  $4000$\\ 
${\tilde d_1}$ & $3400$ &  $3700$\\\hline
\end{tabular}
\end{center}
\end{table}

It should be noted that the focus point is rather sensitive to the (measured) values of the dimensionless MSSM parameters, in particular to the gauge couplings. Changing the unified gauge coupling by only a few percent would lead to other messenger numbers being preferred for the cancellation in Eq.~\eqref{delmz}.

\section{Cosmology and collider phenomenology}
The lightest superparticle in our models can naturally be the gravitino, whose abundance makes it a good dark matter candidate for high reheating temperatures (as also required by thermal leptogenesis) \cite{Bolz:1998ek}. The $\chi^0_1$ higgsino is the NLSP. Its relic density is exceptionally small because of the tiny the $\chi^\pm_1$--$\chi^0_1$ mass splitting, leading to efficient chargino coannihilation. This alleviates the BBN problem which one usually faces with gravitino dark matter and a higgsino NLSP. 

Should the Higgs boson turn out to have a mass below about 120 GeV, evidence for our models should be found by the standard SUSY searches at the LHC. Using dedicated cuts, it may even be possible to discriminate between our models and more generic SUSY scenarios \cite{Bobrovskyi:2011jj}.

On the other hand, if the evidence for a Higgs around 125 GeV solidifies, the soft terms in our model would have to be in the multi-TeV range. While the naturalness problem may be ameliorated by the focus point mechanism, as explained in Section \ref{focsec}, such heavy superparticle masses would make our models very difficult to test. The only kinematically accessible states would be the light higgsinos $\chi^0_{1,2}$ and $\chi^\pm_1$ (in some cases, the $\tilde \tau_1$ can also be sub-TeV). Because of the small mass splittings, decays from one higgsino into another would be near-impossible to detect since they would only give rise to extremely soft jets or leptons \cite{Baer:2011ec,Bobrovskyi:2011jj}.

In that case, the most promising channel to constrain our models at the LHC may be searches for monojets or monophotons, where higgsinos are pair-produced in electroweak processes together with initial-state radiation. The higgsinos invisibly decay into $\chi^0_1$ which leaves the detector, and a single jet or photon plus missing energy is detected. Deciphering the details of the higgsino spectrum will however require a linear collider.

\section*{Acknowledgements}
The author thanks S.~Bobrovskyi, W.~Buchm\"uller, and J.~Hajer for enjoyable collaboration.

\end{document}